%% file: paper.tex
\documentclass[letterpaper,compsoc,twoside]{IEEEtran}
\usepackage{fixltx2e} 
\usepackage{cmap} 
\usepackage{ifthen}
\usepackage[T1]{fontenc}
\usepackage[utf8]{inputenc}
\usepackage{amsmath}

\usepackage[font={small,it},labelfont=bf]{caption}
\usepackage{float}

\pdfoutput=1
\usepackage{scipy}
\makeatletter
\def\PY@reset{\let\PY@it=\relax \let\PY@bf=\relax%
    \let\PY@ul=\relax \let\PY@tc=\relax%
    \let\PY@bc=\relax \let\PY@ff=\relax}
\def\PY@tok#1{\csname PY@tok@#1\endcsname}
\def\PY@toks#1+{\ifx\relax#1\empty\else%
    \PY@tok{#1}\expandafter\PY@toks\fi}
\def\PY@do#1{\PY@bc{\PY@tc{\PY@ul{%
    \PY@it{\PY@bf{\PY@ff{#1}}}}}}}
\def\PY#1#2{\PY@reset\PY@toks#1+\relax+\PY@do{#2}}

\expandafter\def\csname PY@tok@gd\endcsname{\def\PY@tc##1{\textcolor[rgb]{0.63,0.00,0.00}{##1}}}
\expandafter\def\csname PY@tok@gu\endcsname{\let\PY@bf=\textbf\def\PY@tc##1{\textcolor[rgb]{0.50,0.00,0.50}{##1}}}
\expandafter\def\csname PY@tok@gt\endcsname{\def\PY@tc##1{\textcolor[rgb]{0.00,0.27,0.87}{##1}}}
\expandafter\def\csname PY@tok@gs\endcsname{\let\PY@bf=\textbf}
\expandafter\def\csname PY@tok@gr\endcsname{\def\PY@tc##1{\textcolor[rgb]{1.00,0.00,0.00}{##1}}}
\expandafter\def\csname PY@tok@cm\endcsname{\let\PY@it=\textit\def\PY@tc##1{\textcolor[rgb]{0.25,0.50,0.56}{##1}}}
\expandafter\def\csname PY@tok@vg\endcsname{\def\PY@tc##1{\textcolor[rgb]{0.73,0.38,0.84}{##1}}}
\expandafter\def\csname PY@tok@vi\endcsname{\def\PY@tc##1{\textcolor[rgb]{0.73,0.38,0.84}{##1}}}
\expandafter\def\csname PY@tok@mh\endcsname{\def\PY@tc##1{\textcolor[rgb]{0.13,0.50,0.31}{##1}}}
\expandafter\def\csname PY@tok@cs\endcsname{\def\PY@tc##1{\textcolor[rgb]{0.25,0.50,0.56}{##1}}\def\PY@bc##1{\setlength{\fboxsep}{0pt}\colorbox[rgb]{1.00,0.94,0.94}{\strut ##1}}}
\expandafter\def\csname PY@tok@ge\endcsname{\let\PY@it=\textit}
\expandafter\def\csname PY@tok@vc\endcsname{\def\PY@tc##1{\textcolor[rgb]{0.73,0.38,0.84}{##1}}}
\expandafter\def\csname PY@tok@il\endcsname{\def\PY@tc##1{\textcolor[rgb]{0.13,0.50,0.31}{##1}}}
\expandafter\def\csname PY@tok@go\endcsname{\def\PY@tc##1{\textcolor[rgb]{0.20,0.20,0.20}{##1}}}
\expandafter\def\csname PY@tok@cp\endcsname{\def\PY@tc##1{\textcolor[rgb]{0.00,0.44,0.13}{##1}}}
\expandafter\def\csname PY@tok@gi\endcsname{\def\PY@tc##1{\textcolor[rgb]{0.00,0.63,0.00}{##1}}}
\expandafter\def\csname PY@tok@gh\endcsname{\let\PY@bf=\textbf\def\PY@tc##1{\textcolor[rgb]{0.00,0.00,0.50}{##1}}}
\expandafter\def\csname PY@tok@ni\endcsname{\let\PY@bf=\textbf\def\PY@tc##1{\textcolor[rgb]{0.84,0.33,0.22}{##1}}}
\expandafter\def\csname PY@tok@nl\endcsname{\let\PY@bf=\textbf\def\PY@tc##1{\textcolor[rgb]{0.00,0.13,0.44}{##1}}}
\expandafter\def\csname PY@tok@nn\endcsname{\let\PY@bf=\textbf\def\PY@tc##1{\textcolor[rgb]{0.05,0.52,0.71}{##1}}}
\expandafter\def\csname PY@tok@no\endcsname{\def\PY@tc##1{\textcolor[rgb]{0.38,0.68,0.84}{##1}}}
\expandafter\def\csname PY@tok@na\endcsname{\def\PY@tc##1{\textcolor[rgb]{0.25,0.44,0.63}{##1}}}
\expandafter\def\csname PY@tok@nb\endcsname{\def\PY@tc##1{\textcolor[rgb]{0.00,0.44,0.13}{##1}}}
\expandafter\def\csname PY@tok@nc\endcsname{\let\PY@bf=\textbf\def\PY@tc##1{\textcolor[rgb]{0.05,0.52,0.71}{##1}}}
\expandafter\def\csname PY@tok@nd\endcsname{\let\PY@bf=\textbf\def\PY@tc##1{\textcolor[rgb]{0.33,0.33,0.33}{##1}}}
\expandafter\def\csname PY@tok@ne\endcsname{\def\PY@tc##1{\textcolor[rgb]{0.00,0.44,0.13}{##1}}}
\expandafter\def\csname PY@tok@nf\endcsname{\def\PY@tc##1{\textcolor[rgb]{0.02,0.16,0.49}{##1}}}
\expandafter\def\csname PY@tok@si\endcsname{\let\PY@it=\textit\def\PY@tc##1{\textcolor[rgb]{0.44,0.63,0.82}{##1}}}
\expandafter\def\csname PY@tok@s2\endcsname{\def\PY@tc##1{\textcolor[rgb]{0.25,0.44,0.63}{##1}}}
\expandafter\def\csname PY@tok@nt\endcsname{\let\PY@bf=\textbf\def\PY@tc##1{\textcolor[rgb]{0.02,0.16,0.45}{##1}}}
\expandafter\def\csname PY@tok@nv\endcsname{\def\PY@tc##1{\textcolor[rgb]{0.73,0.38,0.84}{##1}}}
\expandafter\def\csname PY@tok@s1\endcsname{\def\PY@tc##1{\textcolor[rgb]{0.25,0.44,0.63}{##1}}}
\expandafter\def\csname PY@tok@ch\endcsname{\let\PY@it=\textit\def\PY@tc##1{\textcolor[rgb]{0.25,0.50,0.56}{##1}}}
\expandafter\def\csname PY@tok@m\endcsname{\def\PY@tc##1{\textcolor[rgb]{0.13,0.50,0.31}{##1}}}
\expandafter\def\csname PY@tok@gp\endcsname{\let\PY@bf=\textbf\def\PY@tc##1{\textcolor[rgb]{0.78,0.36,0.04}{##1}}}
\expandafter\def\csname PY@tok@sh\endcsname{\def\PY@tc##1{\textcolor[rgb]{0.25,0.44,0.63}{##1}}}
\expandafter\def\csname PY@tok@ow\endcsname{\let\PY@bf=\textbf\def\PY@tc##1{\textcolor[rgb]{0.00,0.44,0.13}{##1}}}
\expandafter\def\csname PY@tok@sx\endcsname{\def\PY@tc##1{\textcolor[rgb]{0.78,0.36,0.04}{##1}}}
\expandafter\def\csname PY@tok@bp\endcsname{\def\PY@tc##1{\textcolor[rgb]{0.00,0.44,0.13}{##1}}}
\expandafter\def\csname PY@tok@c1\endcsname{\let\PY@it=\textit\def\PY@tc##1{\textcolor[rgb]{0.25,0.50,0.56}{##1}}}
\expandafter\def\csname PY@tok@o\endcsname{\def\PY@tc##1{\textcolor[rgb]{0.40,0.40,0.40}{##1}}}
\expandafter\def\csname PY@tok@kc\endcsname{\let\PY@bf=\textbf\def\PY@tc##1{\textcolor[rgb]{0.00,0.44,0.13}{##1}}}
\expandafter\def\csname PY@tok@c\endcsname{\let\PY@it=\textit\def\PY@tc##1{\textcolor[rgb]{0.25,0.50,0.56}{##1}}}
\expandafter\def\csname PY@tok@mf\endcsname{\def\PY@tc##1{\textcolor[rgb]{0.13,0.50,0.31}{##1}}}
\expandafter\def\csname PY@tok@err\endcsname{\def\PY@bc##1{\setlength{\fboxsep}{0pt}\fcolorbox[rgb]{1.00,0.00,0.00}{1,1,1}{\strut ##1}}}
\expandafter\def\csname PY@tok@mb\endcsname{\def\PY@tc##1{\textcolor[rgb]{0.13,0.50,0.31}{##1}}}
\expandafter\def\csname PY@tok@ss\endcsname{\def\PY@tc##1{\textcolor[rgb]{0.32,0.47,0.09}{##1}}}
\expandafter\def\csname PY@tok@sr\endcsname{\def\PY@tc##1{\textcolor[rgb]{0.14,0.33,0.53}{##1}}}
\expandafter\def\csname PY@tok@mo\endcsname{\def\PY@tc##1{\textcolor[rgb]{0.13,0.50,0.31}{##1}}}
\expandafter\def\csname PY@tok@kd\endcsname{\let\PY@bf=\textbf\def\PY@tc##1{\textcolor[rgb]{0.00,0.44,0.13}{##1}}}
\expandafter\def\csname PY@tok@mi\endcsname{\def\PY@tc##1{\textcolor[rgb]{0.13,0.50,0.31}{##1}}}
\expandafter\def\csname PY@tok@kn\endcsname{\let\PY@bf=\textbf\def\PY@tc##1{\textcolor[rgb]{0.00,0.44,0.13}{##1}}}
\expandafter\def\csname PY@tok@cpf\endcsname{\let\PY@it=\textit\def\PY@tc##1{\textcolor[rgb]{0.25,0.50,0.56}{##1}}}
\expandafter\def\csname PY@tok@kr\endcsname{\let\PY@bf=\textbf\def\PY@tc##1{\textcolor[rgb]{0.00,0.44,0.13}{##1}}}
\expandafter\def\csname PY@tok@s\endcsname{\def\PY@tc##1{\textcolor[rgb]{0.25,0.44,0.63}{##1}}}
\expandafter\def\csname PY@tok@kp\endcsname{\def\PY@tc##1{\textcolor[rgb]{0.00,0.44,0.13}{##1}}}
\expandafter\def\csname PY@tok@w\endcsname{\def\PY@tc##1{\textcolor[rgb]{0.73,0.73,0.73}{##1}}}
\expandafter\def\csname PY@tok@kt\endcsname{\def\PY@tc##1{\textcolor[rgb]{0.56,0.13,0.00}{##1}}}
\expandafter\def\csname PY@tok@sc\endcsname{\def\PY@tc##1{\textcolor[rgb]{0.25,0.44,0.63}{##1}}}
\expandafter\def\csname PY@tok@sb\endcsname{\def\PY@tc##1{\textcolor[rgb]{0.25,0.44,0.63}{##1}}}
\expandafter\def\csname PY@tok@k\endcsname{\let\PY@bf=\textbf\def\PY@tc##1{\textcolor[rgb]{0.00,0.44,0.13}{##1}}}
\expandafter\def\csname PY@tok@se\endcsname{\let\PY@bf=\textbf\def\PY@tc##1{\textcolor[rgb]{0.25,0.44,0.63}{##1}}}
\expandafter\def\csname PY@tok@sd\endcsname{\let\PY@it=\textit\def\PY@tc##1{\textcolor[rgb]{0.25,0.44,0.63}{##1}}}

\def\PYZus{\char`\_}

\def\PYZdq{\char`\"}

\makeatother

\providecommand*{\DUrole}[2]{%
  \ifcsname DUrole#1\endcsname%
    \csname DUrole#1\endcsname{#2}%
  \else
    \ifcsname docutilsrole#1\endcsname%
      \csname docutilsrole#1\endcsname{#2}%
    \else%
      #2%
    \fi%
  \fi%
}

\ifthenelse{\isundefined{\hypersetup}}{
  \usepackage[colorlinks=true,linkcolor=blue,urlcolor=blue]{hyperref}
  \urlstyle{same} 
}{}

\begin{document}
\newcounter{footnotecounter}\title{Probabilistic Programming and PyMC3}\author{Peadar Coyle$^{\setcounter{footnotecounter}{1}\fnsymbol{footnotecounter}\setcounter{footnotecounter}{2}\fnsymbol{footnotecounter}}$%
          \setcounter{footnotecounter}{1}\thanks{\fnsymbol{footnotecounter} %
          Corresponding author: \protect\href{mailto:peadarcoyle@googlemail.com}{peadarcoyle@googlemail.com}}\setcounter{footnotecounter}{2}\thanks{\fnsymbol{footnotecounter} 9 Rue du Canal Esch-sur-Alzette}\thanks{%

          \noindent%
          Copyright\,\copyright\,2015 Peadar Coyle. This is an open-access article distributed under the terms of the Creative Commons Attribution License, which permits unrestricted use, distribution, and reproduction in any medium, provided the original author and source are credited. http://creativecommons.org/licenses/by/3.0/%
        }}\maketitle
          \renewcommand{\leftmark}{PROC. OF THE 8th EUR. CONF. ON PYTHON IN SCIENCE (EUROSCIPY 2015)}
          \renewcommand{\rightmark}{PROBABILISTIC PROGRAMMING AND PYMC3}

\InputIfFileExists{page_numbers.tex}{}{}
\newcommand*{\docutilsroleref}{\ref}
\newcommand*{\docutilsrolelabel}{\label}
\AtEndDocument{\cleardoublepage}
\begin{abstract}In recent years sports analytics has gotten more and more
popular. We propose a model for Rugby data - in
particular to model the 2014 Six Nations tournament.
We propose a Bayesian hierarchical model to estimate the characteristics that bring a team to lose or win a game, and predict the score of particular matches.

This is intended to be a brief introduction to Probabilistic Programming in Python and in particular the powerful library called PyMC3.\end{abstract}

\section{Introduction%
  \label{introduction}%
}

Probabilistic Programming or Bayesian Statistics \cite{DoingBayes} is what some call a new paradigm.
The aim of this paper is to introduce a Hierarchical model for Rugby Prediction, and also provide an
introduction to PyMC3. Readers who are unfamiliar with Hierarchical models are advised to either read a more thorough exposition online or turn to the excellent textbook on
multilevel modelling \cite{Multilevel}.

Since I am a rugby fan I decide to apply the results of the paper Bayesian Football to the Six Nations. Rugby union is a contact sport that consists of two teams of fifteen players. The objective is to obtain more points than the opposition
through scoring tries or kicking goals over eighty minutes of playing time. Play is started with one team drop kicking the ball from the halfway line towards the opposition.
The rugby ball can be moved up the field by either carrying it or kicking it. However, when passing the ball it can only be thrown laterally or backward. The opposition can
stop players moving up the field by tackling them. Only players carrying the ball can be tackled and once a tackle is completed the opposition can compete for the ball. Play
continues until a try is scored, the ball crosses the side line or dead-ball line, or an infringement occurs. After a team scores points, the non-scoring team restarts the
game at the halfway with a drop kick towards the opposition. The team with the most points at the end wins the game.

Within the Bayesian framework, which naturally accommodates hierarchical models \cite{DoingBayes},  we use here the result proved in {[}Biao{]}\_that assuming two conditionally
independent Poisson variables for the number of points scored, correlation is taken into account, since the observable variables are mixed at an upper level.
Moreover, since we are employing a Bayesian framework, the prediction of the outcome of a new game under the model is provided by the posterior predictive distribution
This predictive distribution is approximated by a Monte Carlo method.

\section{Model%
  \label{model}%
}

My model is based upon the model proposed in \cite{Biao} this is a Hierarchical model for inferring the \emph{strength} of each rugby team in the Six Nations from the data we have about scoring intensity.

Let me introduce some data which we'll need for the model.\begin{Verbatim}[commandchars=\\\{\},fontsize=\footnotesize]
\PY{n}{data\PYZus{}csv} \PY{o}{=} \PY{n}{StringIO}\PY{p}{(}
\PY{l+s+sd}{\PYZdq{}\PYZdq{}\PYZdq{}home\PYZus{}team,away\PYZus{}team,home\PYZus{}score,away\PYZus{}score}
\PY{l+s+sd}{Wales,Italy,23,15}
\PY{l+s+sd}{France,England,26,24}
\PY{l+s+sd}{Ireland,Scotland,28,6}
\PY{l+s+sd}{Ireland,Wales,26,3}
\PY{l+s+sd}{Scotland,England,0,20}
\PY{l+s+sd}{France,Italy,30,10}
\PY{l+s+sd}{Wales,France,27,6}
\PY{l+s+sd}{Italy,Scotland,20,21}
\PY{l+s+sd}{England,Ireland,13,10}
\PY{l+s+sd}{Ireland,Italy,46,7}
\PY{l+s+sd}{Scotland,France,17,19}
\PY{l+s+sd}{England,Wales,29,18}
\PY{l+s+sd}{Italy,England,11,52}
\PY{l+s+sd}{Wales,Scotland,51,3}
\PY{l+s+sd}{France,Ireland,20,22\PYZdq{}\PYZdq{}\PYZdq{}}\PY{p}{)}
\end{Verbatim}
One of the strengths of probabilistic programming is the
ability to infer latent parameters.
These are parameters that can't be measured directly. Our
latent parameter is the strength of each team. We also want
to use this to rank the teams.
The league is made up by a total of T= 6 teams, playing each
other once in a season.

We indicate the number of points scored by the home and
the away team in the g-th game of the season (15 games)
$y_{g1}$ and $y_{g2}$  respectively.

The vector of observed counts $y = (y_{g1}, y_{g2})$
is modelled as independent Poisson:
$y_{gi}| \theta_{gj} =  Poisson(\theta_{gj})$
where the theta parameters represent the scoring intensity
in the g-th game for the team playing at home (j=1) and away
(j=2), respectively.
We model these parameters according to a formulation that
has been used widely in the statistical literature, assuming
a log-linear random effect model

$\log \theta_{g1} = home + att_{h(g)} + def_{a(g)}$

$\log \theta_{g2} = att_{a(g)} + def_{h(g)}$
The parameter home represents the advantage for the team
hosting the game and we assume that this effect is constant
for all the teams and throughout the season. The scoring intensity is determined jointly by the attack
and defense ability of the two teams involved, represented
by the parameters att and def, respectively. Conversely, for each t = 1, ..., T, the team-specific
effects are modelled as exchangeable from a common
distribution $att_{t} = Normal(\mu_{att},\tau_{att})$
and $def_{t} = Normal(\mu_{def},\tau_{def})$

Team strength is decomposed into attacking and defending strength components. A negative defense parameter will sap the mojo from the opposing team's attacking parameter.
I made two tweaks to the model. It didn't make sense to me
to have a $\mu_{att}$ when we're enforcing the sum-to-zero
constraint by subtracting the mean anyway. Also because of the sum-to-zero constraint, it seemed to me that we needed an intercept term in the log-linear model, capturing the average points scored per game by the away team. This we model with the following hyperprior.
$intercept = Normal(0,0.001)$
The hyper-priors on the attack and defense parameters are
also flat $\mu_{att} =   Normal(0,0.001)$,
$\mu_{def}  = Normal(0,0.001)$, $\tau_{att} = \Gamma(0.1,0.1)$ and $\tau_{def} = \Gamma(0.1,0.1)$

\section{Building and executing the model%
  \label{building-and-executing-the-model}%
}

You can see the full code at \cite{Peadar} but the important thing to note that in \cite{PyMC3} the model is all contained in a context manager.
I specified the model and the likelihood function. Fundamentally the Bayesian approach is about calculating posterior distributions. A conventional way to do this is to use a Monte Carlo sampler of which there are many see \cite{DoingBayes}.
I chose the No U Turn Sampler \cite{NUTS} which is a modern sampler for this problem, and we used the Maximum A Posteriori algorithm to find the starting point for that sampler. The Maximum A Posteriori algorithm is a modern optimization approach to finding the starting point. Since convergence of samplers is strongly affected by which starting point is chosen. It is beyong the scope of this article to go into the technicalities but I recommend the following references references as a starting point \cite{NUTS}, \cite{DoingBayes} and {[}PyMC3{]}\_and the references included in those articles.

\section{Results%
  \label{results}%
}

We can use the model above to help us estimate the different distributions of attacking strength and defensive strength.
These are probabilistic estimates and help us better understand the uncertainty in sports analytics.\begin{figure}[]\noindent\makebox[\columnwidth][c]{\includegraphics[scale=0.50]{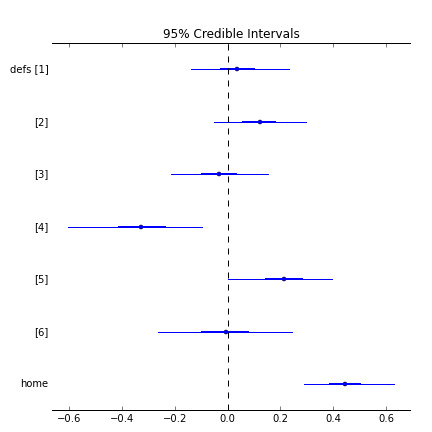}}\end{figure}

Forest plot of the results \DUrole{label}{egfig}

\DUrole{ref}{egfig} is an example of the type of figures that can be generated, which in this example is a forest plot of credible intervals(see \cite{Biao}, and \cite{DoingBayes} for explanations on how to interpret credible intervals)
The estimated ranking of teams is Wales for 1, France for 2, Ireland for 3, Scotland for 4, Italy for 5 and England for 6.

I have built a non-trivial model or generative story for exploring rugby data, I expect that these models can be easily adopted to other sports such as soccer or American
Football. PyMC3 despite being at the time of writing in beta is a useful framework for building Probabilistic Programming models. I was able to show how to use modern MCMC (
Markov Chain Monte Carlo) samplers to approximate a likelihood function (generally one which would be extremely difficult to calculate without numerical methods) and from this
infer latent parameters - that is parameters that are not easy to measure directly. In this case it is \emph{team strength} but there are numerous other applications such as
Stochastic Volatility in Finance \cite{PyMC3}. Also we were able to illustrate how uncertainty estimates such as 'credibility intervals' come out 'for free' from models such as this. I hope that this example and the references inspire you to build your own models and please submit these models to the documentation.

\end{document}

%% file: page_numbers.tex
\setcounter{page}{37}